# Laser-induced ultrafast structural transformations in a thin Fe layer revealed by time-resolved X-ray diffraction

O. Liubchenko[1,*], J. Antonowicz[2], K. Sokolowski-Tinten[3,4], P. Zalden[5], R. Minikayev[1], I. Milov[6, 7, 8], T. J. Albert[3], C. Bressler[5, 9, 10], M. Chojnacki[1], P. Dłużewski[1], P. Dzięgielewski[2], A. Rodriguez-Fernandez[5], K. Fronc[1,11], W. Gawelda[12, 13, 14], K. Georgarakis[15], A.L. Greer[16], I. Jacyna[1], R.W.E. van de Kruijs[8], R. Kamiński[17], D. Khakhulin[5], D. Klinger[1], K. Kosyl[1], K. Kubicek[5, 9, 10], A. Olczak[2], N.T. Panagiotopoulos[16], M. Sikora[18, 19], P. Sun[20], H. Yousef[5], W. Zajkowska-Pietrzak[1] and R. Sobierajski[1,#]

[1]*Institute of Physics, Polish Academy of Sciences, Aleja Lotnikow 32/46, PL-02668 Warsaw, Poland*
[2]*Faculty of Physics, Warsaw University of Technology, Koszykowa 75, 00-662 Warsaw, Poland*
[3]*Faculty of Physics, University of Duisburg-Essen, Lotharstrasse 1, 47048 Duisburg, Germany*
[4]*Center for Nanointegration Duisburg-Essen (CENIDE), University of Duisburg-Essen, Lotharstrasse 1, 47048 Duisburg, Germany*
[5]*European XFEL, Holzkoppel 4, 22869 Schenefeld, Germany*
[6]*Advanced Research Center for Nanolithography (ARCNL), Science Park 106, 1098 XG Amsterdam, the Netherlands*
[7]*Center for Free-Electron Laser Science CFEL, Deutsches Elektronen-Synchrotron DESY, Notkestr. 85, 22607 Hamburg, Germany*
[8]*Industrial Focus Group XUV Optics, MESA+Institute for Nanotechnology, University of Twente, Drienerlolaan 5, 7522 NB Enschede, the Netherlands*
[9]*Department of Physics, Universitat Hamburg, Luruper Chaussee 149, 22761 Hamburg, Germany*
[10]*The Hamburg Centre for Ultrafast Imaging, Luruper Chaussee 149, 22761 Hamburg, Germany*
[11]*Research Centre MagTop, Institute of Physics, Polish Academy of Sciences, Aleja Lotnikow 32/46, PL-02668 Warsaw, Poland*
[12]*Department of Chemistry, Universidad Autónoma de Madrid, Ciudad Universitaria de Cantoblanco 28049 Madrid, Spain*
[13]*IMDEA Nanociencia, Calle Faraday 9, 28049 Madrid, Spain*
[14]*Faculty of Physics, Adam Mickiewicz University, ul. Uniwersytetu Poznańskiego 2, 61-614 Poznan, Poland*
[15]*School of Aerospace, Transport and Manufacturing, Cranfield University, Cranfield, MK43 0AL, UK*
[16]*Department of Materials Science & Metallurgy, University of Cambridge, Cambridge, CB3 0FS, UK*
[17]*Department of Chemistry, University of Warsaw, Żwirki i Wigury 101, 02-089 Warsaw, Poland*
[18]*Academic Centre for Materials and Nanotechnology, AGH University of Krakow, Al. A. Mickiewicza 30, 30-059 Krakow, Poland*
[19]*National Synchrotron Radiation Centre SOLARIS, Jagiellonian University, ul. Czerwone Maki 98, 30-392 Krakow, Poland*
[20]*Dipartimento di Fisica e Astronomia "Galileo Galilei", Universita degli Studi di Padova, Padova 35131, Italy*

*Contact author:liubchenko@ifpan.edu.pl

#Contact author:ryszard.sobierajski@ifpan.edu.pl

## Abstract

The ultrafast structural response of a thin iron film to sub-ps pulsed laser-induced heating has been investigated using time-resolved X-ray diffraction in the partial melting regime. A tetragonal distortion of the bcc-phase emerges at ~6 ps. Its formation is delayed relative to the initial heating (1-2 ps) and partial melting (2-5 ps) of the material, and controlled by the stress release in the quasi-instantaneously pressurized film. The distortion persists for at least 60 ps indicating the formation of a metastable bct phase between the equilibrium bcc and fcc phases.

## I. Introduction

Iron (Fe) is one of the most abundant heavy elements in the universe, main constituent of earth-like planets [1,2], and of major technological relevance [3,4]. Its properties have been extensively studied in a broad range of temperatures and pressures [1,5,6]. The structural phase transitions (SPTs) between different solid polymorphs of iron from stable α-Fe (bcc lattice) under normal conditions to γ-Fe (fcc), δ-Fe (bcc) and ε-Fe (hcp) have been extensively studied using both experimental and theoretical approaches [7-13], and are still the subject of debate [14-17].

Molecular dynamics simulations have been used to investigate ultrafast structural changes in Fe [7,13,17-19]. In the investigated systems, timescales of SPTs ranged from tens to hundreds of picoseconds, depending on the simulation conditions. However, the results of these simulations are strongly dependent on model parameters with the particular challenge that atomic potentials reported in the literature do not allow for simultaneous studies of melting [7], SPTs [8,18], and changes in magnetic properties of Fe [20]. Moreover, in case of arbitrary designed atomic potentials – independent of specific elements – kinetic pathways between bcc and fcc crystal structures and their dynamics during cooling have been recently reported in [21].

Ab-initio calculations have revealed that phonon softening – driven by changes in the temperature-dependent free energy landscape shaped by magnetic order, electronic entropy and lattice vibrational entropy [22-28] – serves as a dynamic precursor to structural modifications [24,29,30]. In Fe, phonon softening of the transverse $(110)$ acoustic $T1$ mode along the $\Gamma - N$ branch has been experimentally observed [25,30]. According to theoretical predictions, it induces a tetragonal distortion of the bcc lattice, which is considered the initial step in the structural transformations along the Bain pathway during the α-γ SPT [24,30,31].

Impulsive heating by sub-picosecond lasers is a common way to induce rapid changes in the free-energy landscape by quasi-instantaneous electronic excitation and fast heating, enabling investigation of subsequent relaxation processes by ultrafast probing methods. Recent time-resolved experimental studies have examined SPTs in pure iron, tracking the α-γ and α-ε SPTs with sub-nanosecond resolution [1,5,32]. The appearance of γ-Fe and ε-Fe phases was observed within hundreds of picoseconds after excitation. However, in those experiments there was no sensitivity (temporal resolution) to the system evolution over structural paths of the transition. Experiments at shorter time-scales were only performed for other metals (mostly fcc) like Au [33,34], Pd [35,36], Pt [37], Al [38], Ni [39], Ag [40], W [41] (bcc). However, only melting was investigated since SPTs were not observed in these materials, both at ultrashort and equilibrium heating conditions.

Our present studies – ultrafast time-resolved X-ray diffraction (XRD) measurements of transient structural states in a thin nanocrystalline Fe layer induced by fs-laser irradiation – are thus the first to address the SPT at ultrashort time-scales corresponding to the characteristic times of relevant phenomena like electronic excitation, rapid lattice heating, phonon softening and stress relaxation.

## II. Methods

Ultrafast laser heating together with XRD structural characterization of Fe thin films was performed using an optical pump – X-ray probe scheme at the FXE instrument [42] of the European XFEL facility. The sample consisted of a matrix of X-ray transparent windows. These windows were composed of 300 nm thick silicon nitride membranes coated with an approx. 27 nm thick nanocrystalline Fe layer with approx. 5 nm native oxide and capped with a 300 nm silicon oxide layer (See Fig. S1a in Supplemental Material (SM) [43]).

The thin Fe layers were excited using laser pulses with 515 nm wavelength and a pulse duration of 0.85 ps (FWHM). The pulses were focused onto the sample at an incidence angle of 3 deg to a spot of approx. 100 μm diameter (FWHM) [47]. The temperature rise in the metallic film was controlled by adjusting the incident laser fluence in the range from 76 to 280 $mJ/cm^2$. However, for this study we focus on heating with a deposited energy density of approx. 166 $mJ/cm^2$ – representative of the partial melting regime obtained at fluences from 95 to 225 $mJ/cm^2$ (See Fig. S6 in SM [43]).

Structural characterization of the excited samples was performed using a transmission XRD geometry (Fig. S2 in the SM [43]). Femtosecond X-ray pulses with a photon energy of 9.43 keV were used to probe the central part (approx. 10 μm diameter) of the laser-excited area. The delay time between the pump and probe beams was controlled in the range from -5 ps (i.e. the X-ray probe pulse arriving before the optical pump) to 60 ps.

## III. Results

Two-dimensional XRD images were azimuthally integrated using the PyFAI software [48] to provide the scattering intensity $I(q)$ as a function of the momentum transfer $q$. Data analysis was focused on the q-range from 2.3 to 3.7 $Å^{-1}$, where the most intense XRD peaks are located. Within this range, a main diffraction peak (110) of the α-Fe (ICSD 48382) was observed in the as-grown sample. Following laser irradiation, the intensity of the Bragg diffraction peak diminished, and a broad "halo" peak appeared, which is interpreted as scattering from the liquid phase. Over time, the original (110) Bragg peak evolved into two distinct but partially overlapping peaks: one with a q-position close to that of the initial diffraction peak before laser irradiation ("high-q peak") and another at a lower q-position ("low-q peak"). To extract information about the behavior of these three components in the XRD patterns, they were deconvoluted into two (narrow) Pseudo-Voigt peaks representing the crystalline phase and a (broad) Gaussian peak for the liquid (Fig. 1).

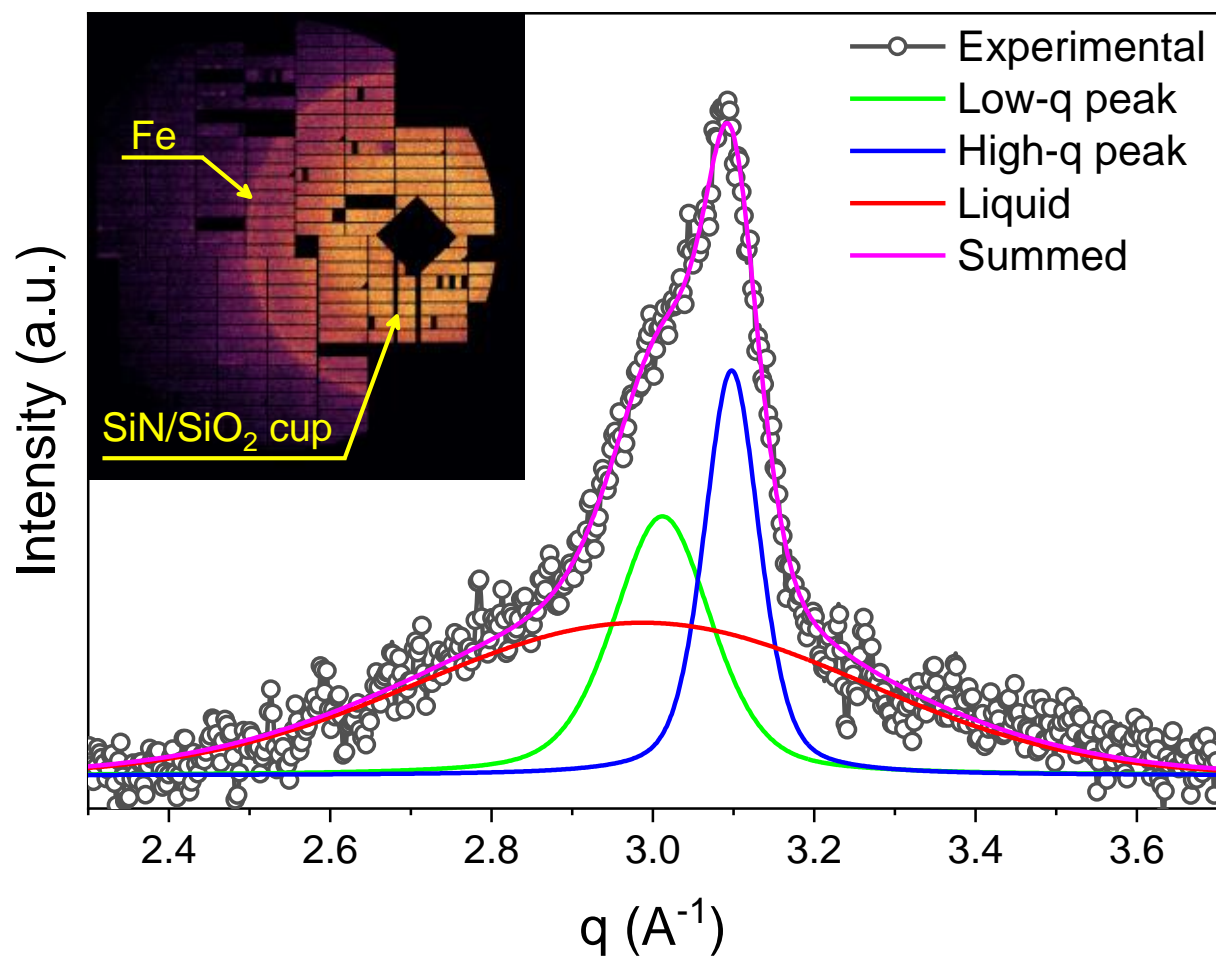

FIG 1 Azimuthally integrated XRD pattern and peak deconvolution for a pump fluence of 166 mJ/cm$^2$ at a delay time 12.5 ps. The inset shows the corresponding 2D detector image, where a continuous Debye-Scherrer ring indicates the presence of a nanocrystalline thin Fe layer. An intense broad ring at lower $q$ values originates from the scattering by the amorphous $SiN/SiO_2$ layers.

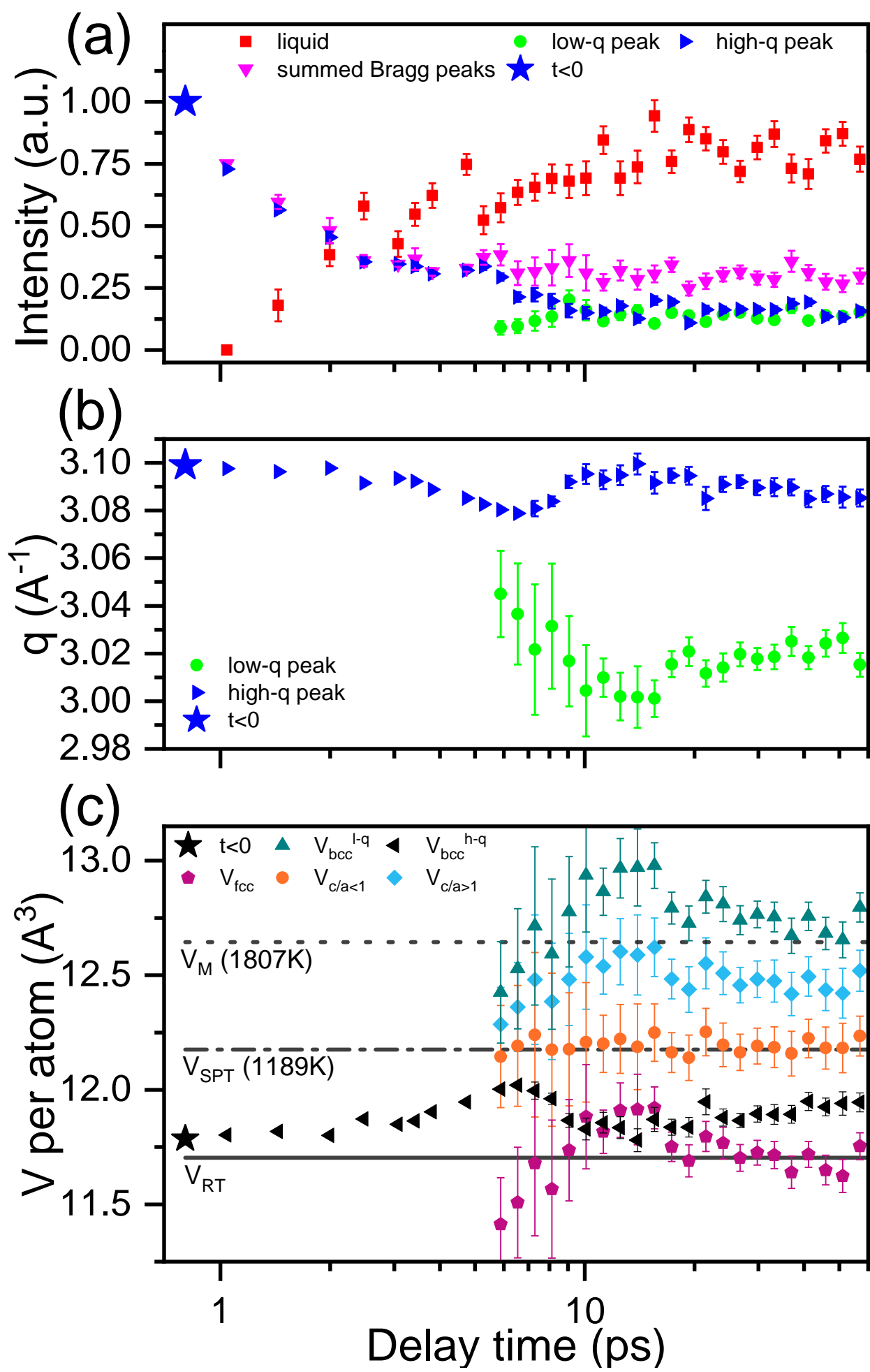

FIG 2 Dependence of peak parameters on delay time for a pump fluence of 166 mJ/cm$^2$: (a) integrated intensity normalized to the signal for the as-grown (solid) and the fully molten state (liquid), (b) Bragg peak positions. (c) Changes in volume per atom $V$ considering the following structural models to explain the observed splitting of the (110) Bragg peak: (I) two bcc phases with different lattice parameters; (II) a mixed fcc and bcc phases (III) tetragonal distortions with $c/a < 1$ or (IV) $c/a > 1$. Literature values of $V$ measured at ambient pressure for bcc α-Fe at room-temperature, near the bcc-fcc SPT temperature (1189K), and bcc δ-Fe near the melting point (1807K) are represented by horizontal lines ($V_{RT}$, $V_{SPT}$, and $V_M$, respectively) [49].

Temporal variations in the fitted peak parameters – including integrated intensities (peak areas) and peak positions – during partial melting of the sample (pump fluence 166 mJ/cm$^2$) are shown in Fig. 2. The time evolution of these parameters for other fluences exhibits similar trends (Fig. S5 in the SM [43]).

The area of the high-q diffraction peak decreases rapidly within the first 5 ps after laser irradiation (Fig. 2a). The broad “halo” emerges at a delay time of $1-2$ ps, with its area increasing first rapidly - accounting for approximately 90% of the total growth – by 5 ps, followed by a slower increase contributing the remaining ~10% of the peak intensity up to 15 ps. The integrated intensity of the “halo” remains almost constant after 15 ps. The low-q peak emerges approximately 6 ps after excitation. Due to the limited ability to deconvolve the XRD signal into two overlapping Bragg peaks, the actual onset of this process may be overestimated. The low-q peak intensity increases in parallel with the decrease in intensity of the high-q peak over delays from approx. 6 to 9 ps. The summed intensity of the Bragg peaks remains nearly unchanged during this interval. For longer delays, up to 15 ps, the Bragg peaks intensity slightly decreases, but beyond 15 ps – up to the maximum studied delay of 60 ps – no significant changes in peak areas are observed.

Before laser irradiation (negative delays), and up to approximately 2 ps, the high-q Bragg peak is centered at 3.1 $Å^{-1}$ (Fig. 2b). It subsequently shifts to its minimum position at $q = 3.08$ $Å^{-1}$ around 6 ps. The low-q peak emerges centered near 3.04 $Å^{-1}$. With increasing delay time, the high-q peak gradually returns to its initial position of 3.1 $Å^{-1}$, while the low-q peak's center continuously shifts to lower values, reaching approx. 3.0 $Å^{-1}$ – resulting in a maximum peak splitting at approx. 12.5 ps. Thereafter, the high-q peak shifts slightly to lower values, settling near 3.09 $Å^{-1}$ at 60 ps, whereas the low-q peak shifts moderately towards higher q, eventually reaching 3.02 $Å^{-1}$.

The dependence of volume per atom on delay time is shown in Fig. 2c. A detailed discussion of this aspect follows in the next section.

## IV. Discussion

### A. Heating and Melting

The initial distribution of the laser-pulse energy decays exponentially with depth, following the Beer-Lambert law. Since the Fe layer is more than twice as thick as the absorption depth (11.7 nm compared to the layer thickness 27 nm), a strong temperature gradient within the electronic subsystem is expected. It drives ultrafast diffusive energy transport, leading to a fast flattening of the temperature depth profile. At the same time, the strong electron-phonon coupling in Fe, as reported in the literature [50,51], results in ultrafast heating of the lattice.

In agreement with these expectations the experimental data indicate that heating of iron occurs almost immediately after laser excitation on a timescale of about 1 ps. The intensity of the α-Fe (110) Bragg peak decreases to approx. 75% of its initial value at room temperature (300 K) within this delay range (Fig. 2a). This agrees with the Debye-Waller factor calculated for this peak at the temperature corresponding to the melting point [52] (See

SM [43]). Despite the heating, almost no shift in the (high-q) Bragg peak position occurs (Fig. 2b), indicating the absence of any lattice expansion at this timescale. The volume per atom, calculated for the bcc phase (See SM [43]), remains near the room temperature level $V_{RT}$. This observation aligns with findings reported in Ref. [38], which similarly noted a delayed lattice expansion relative to the Bragg peaks intensity decrease.

Ultrafast melting of the Fe layer is observed to set in after approx. 1 ps. A broad "halo", interpreted as a signal proportional to the liquid phase content, emerges and grows to about 90% of its final value observed at late time delays (See Fig. 2a and Fig. S5 in SM [43]). Concurrently, the intensity of the high-q Bragg peak continuously decreases, though never reaching zero. This observation evidences the coexistence of the solid and liquid phases, thereby indicating partial melting under the given excitation conditions. The center of the high-q peak shifts from its initial position towards lower q-values, reflecting expansion of the bcc lattice (Fig. 2b). Assuming isotropic 3D expansion of the bcc-phase (mediated by the simultaneous presence of the liquid, see below), the volume per atom increases by 1.4% at delay time of 5 ps.

### B. Tetragonal distortion

For delay times above 6 ps, the diffraction profile develops a clear shoulder and can be best described by the superposition of two peaks, i.e. a low-q and a high-q peak, indicating the formation of a new atomic plane set with different interplanar spacing. Knowledge of the interplanar distances $d_{hkl}$ allows to calculate the lattice parameters and the volume per atom (See Fig. 2c and SM [43] for a calculation details) using the approximation of isotropic expansion of the solid phase component of the partially molten film. This is based on the assumption that the coexistence of the solid and liquid phases establishes hydrostatic conditions although the Fe-layer as a whole can only expand uniaxially normal to the surface. Comparing instead the experimental data to simulations assuming only uniaxial expansion the observed Bragg peak positions at the point of maximum peak splitting require unphysical (negative, in the order of a few GPa) values of the uniaxial stress (See discussion in SM [43]).

We have considered four independent structural models leading to the observed two Bragg peaks (Fig. 2c): (I) Coexistence of two bcc phases with different lattice parameters, caused by a non-uniform temperature or strain distribution. This scenario is deemed unrealistic, since the splitting persists at least up to 60 ps when uniform conditions of temperature and pressure over the whole sample depth can be expected. Moreover, the calculated volume per atom for both of the bcc phases are implausible. $V_{bcc}^{l-q}$ exceeds the maximum value $V_M$ obtained for δ-Fe (high-temperature bcc phase of Fe) near the melting temperature for the case of slow heating at ambient pressure for the one phase and $V_{bcc}^{h-q}$ remains near to the room temperature value $V_{RT}$ for the other. (II) The coexistence of bcc and fcc Fe phases, as suggested by the equilibrium Fe phase diagram, can be excluded as a plausible explanation for the second Bragg peak. The α-γ transformation in Fe should reduce the intensity of the Bragg peak to approximately half [53,54], yet Fig. 2a shows the summed intensity remaining essentially constant in a delay range from approx. $6 - 60$ ps in spite of the increase of the intensity of the low-q peak. (See also Fig. S5 in SM [43]). The calculated volume per atom $V_{bcc}^{h-q}$ and $V_{fcc}$ for both phases remains close or even lower than the $V_{RT}$. Moreover, the $(200)$ Bragg peak from the fcc phase at $q = 3.44$ Å$^{-1}$ is not observed.

The only viable mechanism explaining the experimental data is symmetry breaking via a tetragonal distortion of the bcc unit cell in Fe (formation of bct-Fe). Two regimes have to be distinguished depending on the $c/a$ ratio of the unit-cell parameters [16,26,55]. (III) For $c/a < 1$ the volume per atom calculated from the peak position is comparable with $V_{SPT}$, which appears unlikely for Fe heated up to the melting temperature $T_M$. (IV) For $c/a > 1$, the computed volume per atom aligns with $V_M$ measured under equilibrium conditions near the melting temperature $T_M$. Therefore, a tetragonal distortion with $c/a > 1$ offers the most consistent explanation. Moreover, as a first step of the α-γ SPT in Fe along the Bain path [6,13,16] an initial bct distortion with $c/a > 1$ is expected: the lattice parameter $c$ increases and lattice parameters $a$ and $b$ decrease, which is in full agreement with our observations. The observed intensity ratio $I_{low-q}/I_{high-q}$ is 1:1 and thus deviates from the 2:1 ratio expected for bct-Fe. However, it is worth noting that similar ratios were reported in [56] for as-grown bct-Fe nanoparticles.

In the $c/a > 1$ scenario, the lowering of the unit cell symmetry leads to the splitting of the Bragg peak of family $\{110\}$ of the bcc-phase into two peaks: $(110)$ and $\{(011),(101)\}$ of bct-Fe [53]. The $(110)$ peak of bct-Fe "inherits" the same interplanar spacing $d_{110}^{bct}$ as the original bcc phase $d_{110}^{bcc}$ and is, therefore attributed to the high-q peak. The $\{(011),(101)\}$ peak exhibits a larger interplanar spacing $d_{\{(011),(101)\}}^{bct}$ and corresponds to the observed low-q peak (Fig. 3c). We assume that the (randomly orientated) crystalline structure is uniform over the whole XRD probed volume. Each peak is formed by scattering from a separate set of nanocrystallites whose orientation allows the Bragg condition to be fulfilled, thus both peaks can be measured simultaneously.

Based on the interpretation of the Bragg peaks presented above we propose the following scheme of the structural transformations in the Fe lattice after sub-ps laser excitation (Fig. 3). Heating of the lattice, due to strong electron-phonon coupling, occurs within approx. 1 ps after excitation. This is too fast to allow for a relaxation of the lattice and thermal pressure in the order of 10 GPa (See SM [43]) builds up. Under these high temperature-high pressure conditions the material partially melts, which might also have an effect on the pressure due to the lower density of the liquid. At very early times the remaining solid material (still in the bcc phase) is volumetrically confined and any structural transformation of the solid requiring large volume changes – i.e. other than melting – is suppressed. Only a relatively small (1.4%) 3D lattice expansion occurs under the isotropic conditions mediated by the molten volume surrounding the nanocrystallites (Fig. 3b). Stress is relaxed by uniaxial expansion of the whole Fe layer within a characteristic time $t_{exp} = d/c_s$ ($d$: film thickness; $c_s$: sound velocity). This lifts the restrictions with respect to the volume change of individual crystallites and enables the tetragonal distortion and the associated increase of the volume per atom (Fig. 3c). With $d = 27$ nm and $c_s = 3.8 - 5.9$ nm/ps in the partially molten film (the lower and upper limit represent the sound velocity in liquid [57] and solid Fe [2], respectively), $t_{exp}$ can be estimated as $t_{exp} = 4.6 - 7.1$ ps, in rather good agreement with the observed onset of peak-splitting at around 6 ps. Remarkably, the onset of the tetragonal distortion coincides with the volume per atom reaching the $V_{SPT}$ threshold, obtained under equilibrium conditions when the sample temperature reaches the transition temperature of the α-γ SPT [49] (Fig. 2c). The volume per atom $V_{bct}^{c/a>1}$ (calculated already for bct-Fe) continues to increase up to a delay time of approx. 12.5 ps, when it reaches its maximum value of 12.6 $A^3$ – close to, but not exceeding $V_M$ – the value reported for δ-Fe [49] near the melting temperature. At this moment the tetragonal distortion becomes maximal with $c/a = 1.065$. The tetragonal distortion persists for at least 60 ps with the volume per atom slightly decreasing down to 12.4 $A^3$ at the longest delay measured. No evidence for the formation of γ-Fe or ε-Fe was found within this timescale.

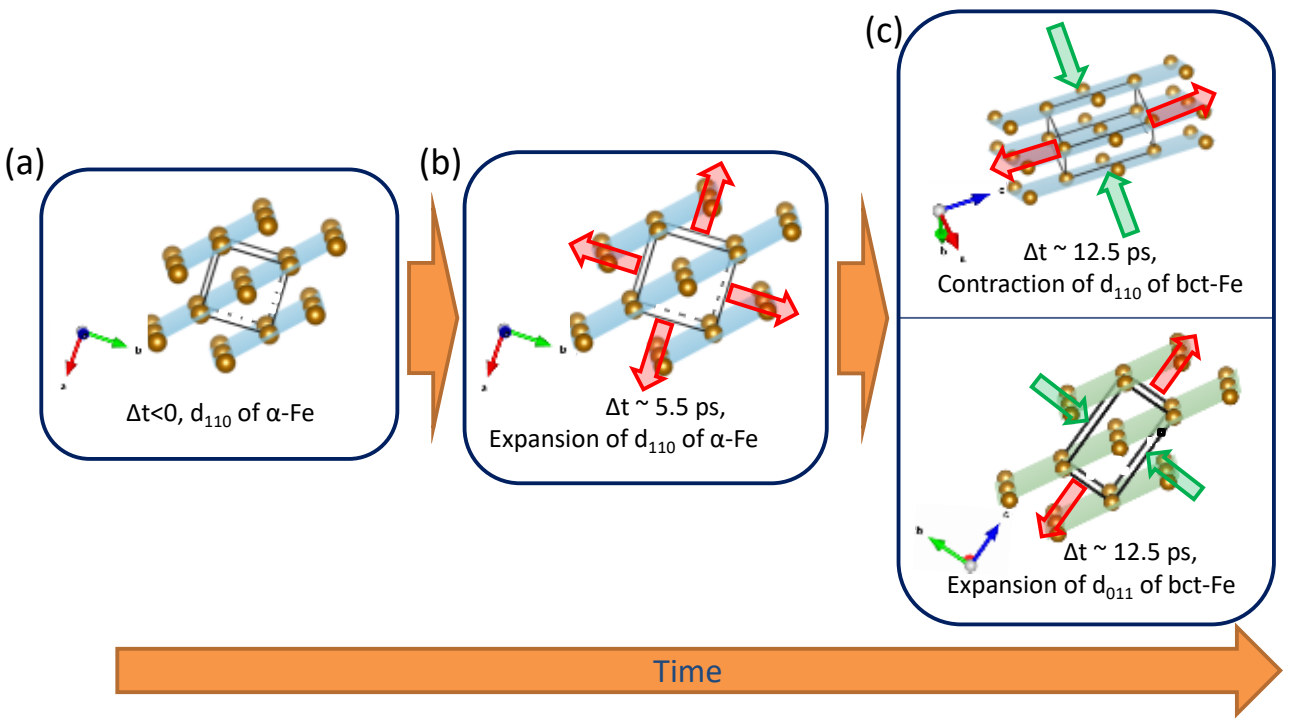

FIG 3 Sketch of changes in interplanar spacing and unit cell distortion for selected delay times [58]: a) “cold” lattice before laser irradiation; b) 3D thermal expansion of the bcc lattice; c) bct distortion and lattice expansion. Solid arrows indicate expansion (red) and contraction (green) of the unit cell along specific directions. Lattice planes, contributing to the observed high-q and low-q Bragg peaks, are marked with blue and green, respectively.

## V. Conclusions

The tetragonal distortion of the bcc lattice in Fe is the initial step of the bcc-fcc (α-γ) SPT, as described by the Bain pathway [13,16]. According to theoretical simulations this transitions is driven by phonon softening [24,27,30]. By short-pulse laser excitation and ultrafast XRD-probing we were able to separate in time the different steps leading ultimately to the long-lived bct-Fe observed in the experiment. In particular, the data reveal that the formation of bct-Fe is delayed relative to “free energy-altering” processes, such as electronic heating and magnetic disordering [59] as well as lattice heating, but also with respect to melting of part of the film. The evolution of the tetragonal distortion is in line with a sound-velocity limited expansion of the film as a whole and thus strongly dependent on the relaxation of the uniaxial stress generated by ultrafast heating of the film.

Notably, the bct distortion persists for at least 60 ps, although for the given pressure/temperature conditions (in the vicinity of the melting threshold), either high temperature bcc or fcc phases are expected according to the equilibrium phase diagram of Fe. This indicates the existence of a local free energy minimum hindering further transformation and formation of a metastable bct phase of Fe, similar as predicted in [21].

The experimental results offer new insight into ultrafast structural dynamics during the early stages of solid-solid phase transitions in Fe and the role of different driving processes – heating, phonon softening and stress release. They highlight the need for further theoretical simulations of structural transformations in materials that go beyond the current static approach, addressing dynamical effects and acknowledging the critical roles of both temperature and stress under fast driving conditions.

**Acknowledgments**

We acknowledge European XFEL in Schenefeld, Germany, for provision of X-ray free-electron laser beamtime at the Scientific Instrument FXE (Femtosecond X-Ray Experiments) and would like to thank the staff for their assistance. The access to the European XFEL was supported by a grant of the Polish Ministry of Education and Science - decision no. 2022/WK/13. This work was supported by the National Science Centre, Poland, grants agreement No 2016/21/D/ST4/03753, No 2017/27/B/ST3/02860 and No 2021/43/B/ST5/02480. K.S.-T. and T.J.A. acknowledge financial support by the Deutsche Forschungsgemeinschaft (DFG, German Research Foundation) through Project No. 278162697-SFB 1242. WG acknowledges funding from the Spanish Ministry of Science, Innovation and Universities (MICIU) through "Proyectos de I+D+I 2022" grant (Ref. PID2022-140257NB-I00) and funding from the "Severo Ochoa" Programme for Centres of Excellence in R&D (CEX2020-001039S/AEI/10.13039/501100011033).

## References

[1] T. Sano et al., *X-Ray Free Electron Laser Observation of Ultrafast Lattice Behaviour under Femtosecond Laser-Driven Shock Compression in Iron*, Sci. Rep. **13**, 13796 (2023).

[2] F. Decremps, D. Antonangeli, M. Gauthier, S. Ayrinhac, M. Morand, G. Marchand, F. Bergame, and J. Philippe, *Sound Velocity of Iron up to 152 GPa by Picosecond Acoustics in Diamond Anvil Cell*, Geophys. Res. Lett. **41**, 1459 (2014).

[3] M. Altay, H. Aydın, and A. Karşı, *Effect of Heat Input on Martensitic Stainless Steel Laser Clad Characteristics on Ductile Cast Iron*, Weld. World **1** (2025).

[4] K. Yu, H. Shi, P. Zhang, Z. Yu, H. Yan, and Q. Lu, *Micro/Nanoengineering of Functionalized Metal Surfaces Based on Short/Ultra-Short-Pulsed Lasers: A Review*, J. Mater. Sci. **59**, 1819 (2024).

[5] H. Hwang et al., *Subnanosecond Phase Transition Dynamics in Laser-Shocked Iron*, Sci. Adv. **6**, 5132 (2020).

[6] R. Fréville, A. Dewaele, N. Bruzy, V. Svitlyk, and G. Garbarino, *Comparison between Mechanisms and Microstructures of α-γ, γ-ϵ, and α-ϵ-α Phase Transitions in Iron*, Phys. Rev. B **107**, 104105 (2023).

[7] K. Li et al., *Determination of the Accuracy and Reliability of Molecular Dynamics Simulations in Estimating the Melting Point of Iron: Roles of Interaction Potentials and Initial System Configurations,* J. Mol. Liq. **290**, 111204 (2019).

[8] X. Ou, *Molecular Dynamics Simulations of Fcc-to-Bcc Transformation in Pure Iron: A Review*, Mater. Sci. Technol. **33**, 822 (2017).

[9] X. F. Zhang and Y. I. Komizo, *In Situ Investigation of the Allotropic Transformation in Iron*, Steel Res. Int. **84**, 751 (2013).

[10] E. Boulard, C. Denoual, A. Dewaele, A. King, Y. Le Godec, and N. Guignot, *Following the Phase Transitions of Iron in 3D with X-Ray Tomography and Diffraction under Extreme Conditions*, Acta Mater. **192**, 30 (2020).

[11] S. Merkel, H. P. Liermann, L. Miyagi, and H. R. Wenk, *In Situ Radial X-Ray Diffraction Study of Texture and Stress during Phase Transformations in Bcc-, Fcc-and Hcp-Iron up to 36 GPa and 1000 K*, Acta Mater. **61**, 5144 (2013).

[12] L. Miyagi, M. Kunz, J. Knight, J. Nasiatka, M. Voltolini, and H. R. Wenk, *In Situ Phase Transformation and Deformation of Iron at High Pressure and Temperature*, J. Appl. Phys. **104**, 103510 (2008).

[13] L. Sandoval, H. M. Urbassek, and P. Entel, *The Bain versus Nishiyama-Wassermann Path in the Martensitic Transformation of Fe*, New J. Phys. **11**, 103027 (2009).

[14] F.-C. Wang, Q.-J. Ye, Y.-C. Zhu, and X.-Z. Li, *Crystal-Structure Matches in Solid-Solid Phase Transitions*, Phys. Rev. Lett. **132**, 086101 (2024).

[15] N. Wang, T. Hammerschmidt, T. Hickel, J. Rogal, and R. Drautz, *Influence of Spin Fluctuations on Structural Phase Transitions of Iron*, Phys. Rev. B **107**, 104108 (2023).

[16] L. H. Zhang, M. J. Cheng, X. H. Shi, J. W. Shuai, and Z. Z. Zhu, *Bain and Nishiyama-Wassermann Transition Path Separation in the Martensitic Transitions of Fe*, RSC Adv. **11**, 3043 (2021).

[17] Q. Li et al., *The Combined Effects of Uniaxial Pressure and Defects on the Mechanism of Martensitic Transformation in Pure Iron: A Molecular Dynamics Study*, Mater. Today Commun. **39**, 109092 (2024).

[18] J. Meiser and H. M. Urbassek, *$\alpha \leftrightarrow \gamma$ Phase Transformation in Iron: Comparative Study of the Influence of the Interatomic Interaction Potential*, Model. Simul. Mater. Sci. Eng. **28**, 055011 (2020).

[19] X. Ou and M. Song, *Deformation Mechanisms of Mechanically Induced Phase Transformations in Iron*, Comput. Mater. Sci. **162**, 12 (2019).

[20] S. L. Dudarev and P. M. Derlet, *A 'Magnetic' Interatomic Potential for Molecular Dynamics Simulations,* J. Phys. Condens. Matter **17**, 7097 (2005).

[21] H. Pan and J. Dshemuchadse, *Kinetic Pathways of Solid–Solid Phase Transitions Dictated by Short-Range Interactions*, Proc. Natl. Acad. Sci. 122, e2507403122 (2025).

[22] Q. Han, T. Birol, and K. Haule, *Phonon Softening Due to Melting of the Ferromagnetic Order in Elemental Iron*, Phys. Rev. Lett. **120**, 187203 (2018).

[23] S. Yang, Y. Wang, Z.-K. Liu, and Y. Zhong, *Ab Initio Studies on Structural and Thermodynamic Properties of Magnetic Fe*, Comput. Mater. Sci. **227**, 112299 (2023).

[24] I. Leonov, A. I. Poteryaev, Y. N. Gornostyrev, A. I. Lichtenstein, M. I. Katsnelson, V. I. Anisimov, and D. Vollhardt, *Electronic Correlations Determine the Phase Stability of Iron up to the Melting Temperature*, Sci. Rep. **4**, 5585 (2014).

[25] J. Neuhaus, W. Petry, and A. Krimmel, *Phonon Softening and Martensitic Transformation in α-Fe*, Phys. B Condens. Matter **234–236**, 897 (1997).

[26] G. Grimvall, B. Magyari-Köpe, V. Ozoliņš, and K. A. Persson, *Lattice Instabilities in Metallic Elements*, Rev. Mod. Phys. **84**, 945 (2012).

[27] Y. Ikeda, A. Seko, A. Togo, and I. Tanaka, *Phonon Softening in Paramagnetic Bcc Fe and Its Relationship to the Pressure-Induced Phase Transition*, Phys. Rev. B **90**, 134106 (2014).

[28] F. Körmann, A. Dick, B. Grabowski, T. Hickel, and J. Neugebauer, *Atomic Forces at Finite Magnetic Temperatures: Phonons in Paramagnetic Iron*, Phys. Rev. B **85**, 125104 (2012).

[29] W. Petry, *Dynamical Precursors of Martensitic Transitions*, Le J. Phys. IV **05**, C2 (1995).

[30] J. Neuhaus, M. Leitner, K. Nicolaus, W. Petry, B. Hennion, and A. Hiess, *Role of Vibrational Entropy in the Stabilization of the High-Temperature Phases of Iron*, Phys. Rev. B **89**, 184302 (2014).

[34] I. Leonov, A. I. Poteryaev, V. I. Anisimov, and D. Vollhardt, *Electronic Correlations at the α-γ Structural Phase Transition in Paramagnetic Iron*, Phys. Rev. Lett. **106**, 106405 (2011).

[32] H. S. Park, O.-H. Kwon, J. S. Baskin, B. Barwick, and A. H. Zewail, *Direct Observation of Martensitic Phase-Transformation Dynamics in Iron by 4D Single-Pulse Electron Microscopy,* Nano Lett. **9**, 3954 (2009).

[33] M. Z. Mo et al., *Heterogeneous to Homogeneous Melting Transition Visualized with Ultrafast Electron Diffraction*, Science. **360**, 1451 (2018).

[34] T. A. Assefa et al., *Ultrafast X-Ray Diffraction Study of Melt-Front Dynamics in Polycrystalline Thin Films*, Sci. Adv. **6**, (2020).

[35] A. F. Suzana et al., *Compressive Effects in Melting of Palladium Thin Films Studied by Ultrafast X-Ray Diffraction*, Phys. Rev. B **107**, 214303 (2023).

[36] J. Antonowicz et al., *Structural Pathways for Ultrafast Melting of Optically Excited Thin Polycrystalline Palladium Films*, Acta Mater. **276**, 120043 (2024).

[37] L. Gelisio et al., *Infrared-Induced Ultrafast Melting of Nanostructured Platinum Films Probed by an x-Ray Free-Electron Laser*, Phys. Rev. B **110**, 144303 (2024).

[38] J. Wu, M. Tang, L. Zhao, P. Zhu, T. Jiang, X. Zou, L. Hong, S.-N. Luo, D. Xiang, and J. Zhang, *Ultrafast Atomic View of Laser-Induced Melting and Breathing Motion of Metallic Liquid Clusters with MeV Ultrafast Electron Diffraction*, Proc. Natl. Acad. Sci. **119**, (2022).

[39] R. Li, H. E. Elsayed-Ali, J. Chen, D. Dhankhar, A. Krishnamoorthi, and P. M. Rentzepis, *Ultrafast Time-Resolved Structural Changes of Thin-Film Ferromagnetic Metal Heated with Femtosecond Optical Pulses*, J. Chem. Phys. **151**, 124702 (2019).

[40] A. O. Er, J. Chen, J. Tang, and P. M. Rentzepis, *Coherent Acoustic Wave Oscillations and Melting on Ag(111) Surface by Time Resolved x-Ray Diffraction*, Appl. Phys. Lett. **100**, 151910 (2012).

[41] H. Zhang, C. Li, E. Bevillon, G. Cheng, J. P. Colombier, and R. Stoian, *Ultrafast Destructuring of Laser-Irradiated Tungsten: Thermal or Nonthermal Process*, Phys. Rev. B **94**, 224103 (2016).

[42] D. Khakhulin et al., *Ultrafast X-Ray Photochemistry at European XFEL: Capabilities of the Femtosecond X-Ray Experiments (FXE) Instrument*, Appl. Sci. **10**, 995 (2020).

[43] See Supplemental Material at [URL will be inserted by publisher] for details of (1) samples, experimental and data processing; (2) additional experimental results; (3) Debye-

Waller factor and $V$ per atom calculation; (4) details of the theoretical models of acoustic waves propagation and thermal stress, which includes Refs. [44–46].

[44] W. Lu, M. Nicoul, U. Shymanovich, F. Brinks, M. Afshari, A. Tarasevitch, D. Von Der Linde, and K. Sokolowski-Tinten, *Acoustic Response of a Laser-Excited Polycrystalline Au-Film Studied by Ultrafast Debye-Scherrer Diffraction at a Table-Top Short-Pulse x-Ray Source*, AIP Adv. **10**, 35015 (2020).

[45] A. Yoneda and A. Kubo, *Simultaneous Determination of Mean Pressure and Deviatoric Stress Based on Numerical Tensor Analysis: A Case Study for Polycrystalline x-Ray Diffraction of Gold Enclosed in a Methanol–Ethanol Mixture*, J. Phys. Condens. Matter **18**, S979 (2006).

[46] M. Gauthier, *Effects of Uniaxial Stress on X-Ray Diffraction Spectra*, High Press. Res. **22**, 779 (2002).

[47] J. Chalupský et al., *Spot Size Characterization of Focused Non-Gaussian X-Ray Laser Beams*, Opt. Express **18**, 27836 (2010).

[48] G. Ashiotis, A. Deschildre, Z. Nawaz, J. P. Wright, D. Karkoulis, F. E. Picca, and J. Kieffer, *The Fast Azimuthal Integration Python Library: PyFAI*, J. Appl. Crystallogr. **48**, 510 (2015).

[49] Z. S. Basinski, W. Hume-Rothery, and A. L. Sutton, *The Lattice Expansion of Iron*, Proc. R. Soc. London. Ser. A. Math. Phys. Sci. **229**, 459 (1955).

[50] N. Medvedev and I. Milov, *Electron-Phonon Coupling in Metals at High Electronic Temperatures*, Phys. Rev. B **102**, 064302 (2020).

[51] N. Rothenbach et al., *Microscopic Nonequilibrium Energy Transfer Dynamics in a Photoexcited Metal/Insulator Heterostructure*, Phys. Rev. B **100**, 174301 (2019).

[52] H. X. Gao and L.-M. Peng, *Parameterization of the Temperature Dependence of the Debye–Waller Factors*, Acta Crystallogr. Sect. A Found. Crystallogr. **55**, 926 (1999).

[53] B. D. Cullty and S. R. Stock, Elements of X-Ray Diffraction: Third Edition (Prentice-Hall, 2001).

[54] A. K. De, D. C. Murdock, M. C. Mataya, J. G. Speer, and D. K. Matlock, *Quantitative Measurement of Deformation-Induced Martensite in 304 Stainless Steel by X-Ray Diffraction*, Scr. Mater. **50**, 1445 (2004).

[55] I. K. Razumov, Y. N. Gornostyrev, and M. I. Katsnelson, *Towards the Ab Initio Based Theory of Phase Transformations in Iron and Steel*, Phys. Met. Metallogr. **118**, 362 (2017).

[56] J. Liu, K. Schliep, S. H. He, B. Ma, Y. Jing, D. J. Flannigan, and J. P. Wang, *Iron Nanoparticles with Tunable Tetragonal Structure and Magnetic Properties*, Phys. Rev. Mater. **2**, 054415 (2018).

[57] P. M. Nasch, M. H. Manghnani, and R. A. Secco, *Sound Velocity Measurements in Liquid Iron by Ultrasonic Interferometry*, J. Geophys. Res. Solid Earth **99**, 4285 (1994).

[58] K. Momma and F. Izumi, *VESTA 3 for Three-Dimensional Visualization of Crystal, Volumetric and Morphology Data*, J. Appl. Crystallogr. **44**, 1272 (2011).

[59] E. Carpene, E. Mancini, C. Dallera, M. Brenna, E. Puppin, and S. De Silvestri, *Dynamics of Electron-Magnon Interaction and Ultrafast Demagnetization in Thin Iron Films*, Phys. Rev. B **78**, 174422 (2008).

# Supplemental material to “Laser-induced ultrafast structural transformations in a thin Fe layer revealed by time-resolved X-ray diffraction”

O. Liubchenko[1,*], J. Antonowicz[2], K. Sokolowski-Tinten[3,4], P. Zalden[5], R. Minikayev[1], I. Milov[6, 7, 8], T. J. Albert[3], C. Bressler[5, 9, 10], M. Chojnacki[1], P. Dłużewski[1], P. Dzięgielewski[2], A. Rodriguez-Fernandez[5], K. Fronc[1,11], W. Gawelda[12, 13, 14], K. Georgarakis[15], A.L. Greer[16], I. Jacyna[1], R.W.E. van de Kruijs[8], R. Kamiński[17], D. Khakhulin[5], D. Klinger[1], K. Kosyl[1], K. Kubicek[5, 9, 10], A. Olczak[2], N.T. Panagiotopoulos[16], M. Sikora[18, 19], P. Sun[20], H. Yousef[5], W. Zajkowska-Pietrzak[1] and R. Sobierajski[1,#]

## Experimental

### 1. Samples and experimental setup of the pump-probe experiment

Figure S1 shows a schematic layout of the sample (a) and the cross-sectional TEM micrograph of the as-deposited films (b). The sample consisted of a matrix of X-ray transparent windows, composed of 300 nm-thick silicon nitride membranes coated with an approx. 27 nm-thick nanocrystalline Fe layer, deposited by magnetron sputtering in an Ar atmosphere. The metal layer was capped with a 300 nm-thick silicon oxide layer, with an approx. 5 nm-thick native iron oxide forming at the interface (see Fig. S1b for TEM cross-sectional image of the iron layer).

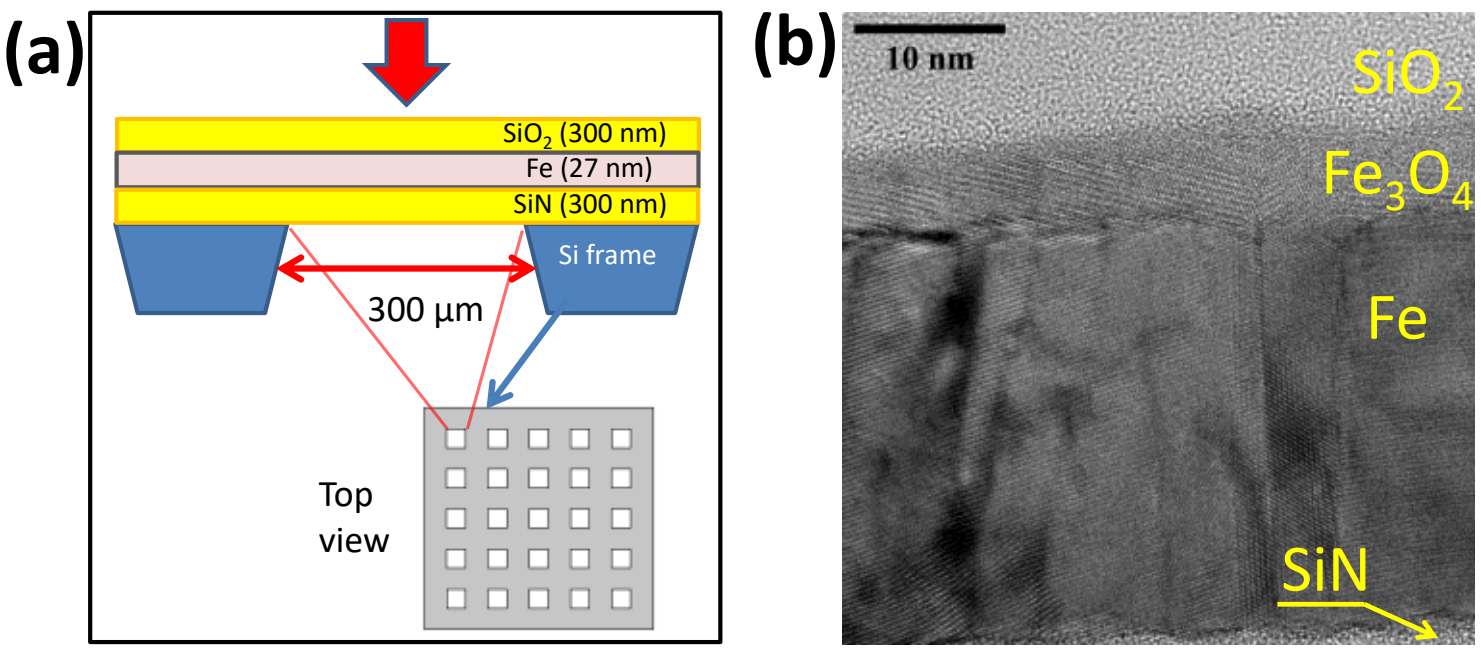

FIG S1 Schematic layout of the sample (a) and cross-sectional TEM of the as-deposited thin Fe film (b).

The experiment was performed at the FXE instrument of the EuXFEL [1]. The experimental set-up is shown in Fig. S2 – see also the supplementary materials in [2]. Thin Fe layers were excited using laser pulses with 515 nm wavelength and a pulse duration of 0.85 ps (FWHM) – see the red arrow in Fig. S2.

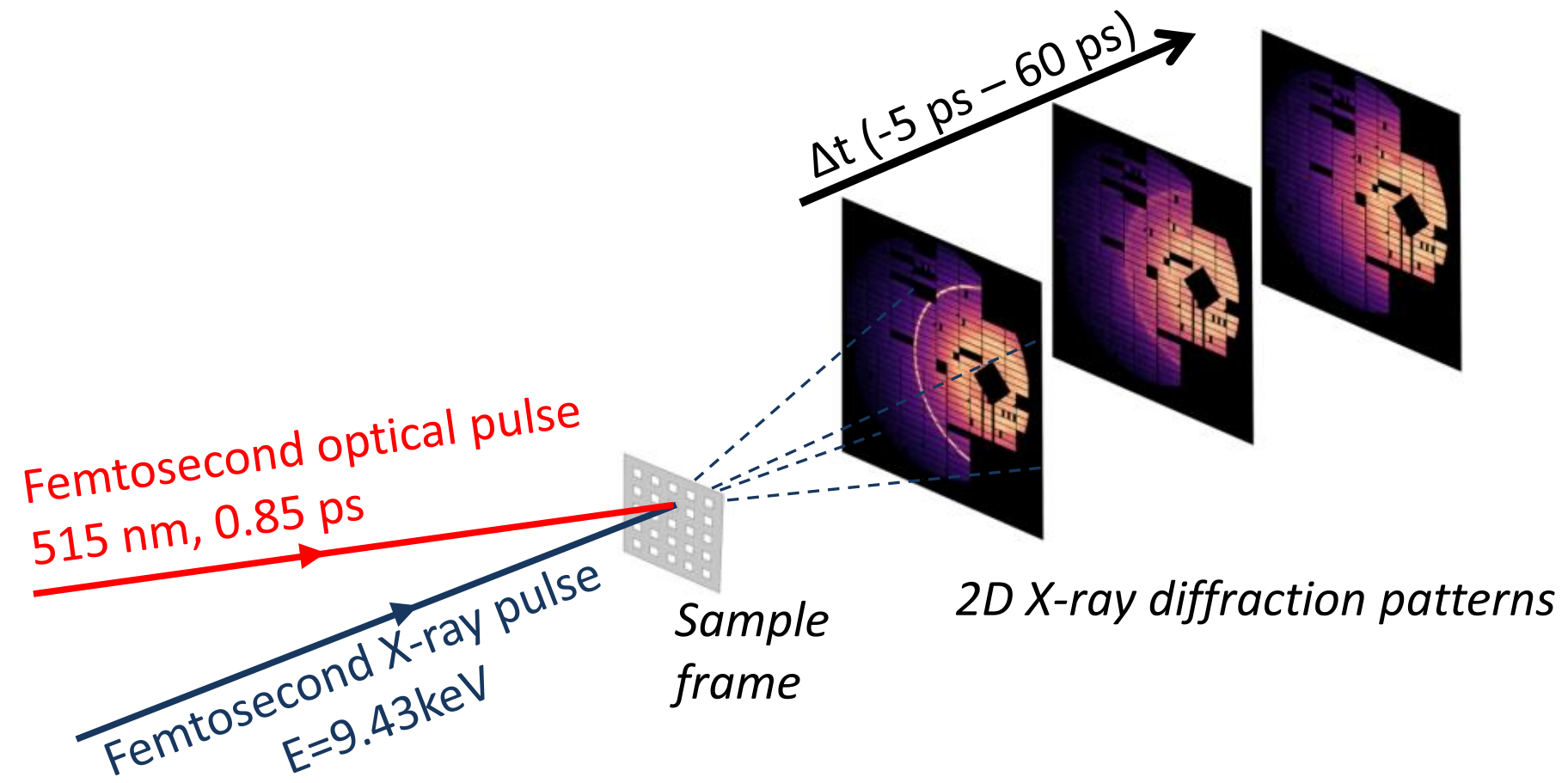

FIG S2. Schematic layout of the pump-probe experiment.

The pulses were focused onto the sample from the silicon oxide side at an incidence angle of 3 deg to a spot of approx. 100 μm diameter (FWHM) [3]. The temperature rise in the metallic film was controlled by adjusting the incidence pump pulse fluence $F_i$ in a range from $\sim 76 - 280$ mJ/cm$^2$. From the reflections/transmission measurements of the our sample the following values: $R = (58 \pm 3)\%$; $T = (4 \pm 1)\%$ were obtained. The resulting absorption value is $A = (38 \pm 4)\%$. Therefore, the fluence absorbed by the Fe layer ranges from $\sim 29 - 106$ mJ/cm$^2$, corresponding to an energy density range of$\sim 1.4 - 5$ MJ/kg (assuming homogeneous distribution over the film).

The sample was irradiated at a repetition rate of approx. 1 Hz in single-pulse mode, i.e. after each exposure to an optical pump pulse and an X-ray probe pulse, the sample was moved, ensuring a “fresh” window in the sample array for the subsequent pump-probe event. Structural characterization of the excited samples was performed using the normal incidence transmission Debye-Scherrer X-ray diffraction geometry. Femtosecond X-ray pulses (blue line in Fig. S2) with a photon energy of 9.43 keV were used to probe the central part (approx. 10 μm diameter) of the laser-excited area. Samples were placed in a vacuum chamber to reduce the background from air-scattered radiation. The delay time between the pump and probe beams was controlled in the range from -5 ps (i.e. the X-ray probe pulse arriving before the optical pump) to 60 ps. The scattered radiation was recorded using the Large Pixel Detector [1] available at FXE. The momentum transfer ($q$) range was constrained between 1 Å$^{-1}$ to 4.5 Å$^{-1}$, determined by the geometry of the experimental setup, including the beamstop blocking the direct X-ray beam and the window of the vacuum chamber.

## 2. XRD data analysis

The detector geometry was calibrated using a $LaB_6$ reference sample (SRM660b) and the PyFAI package [4]. After this, two-dimensional (2D) XRD images were integrated into one-dimensional (1D) patterns with the same package.

*Background subtraction and normalization*

To extract reliable structural information from the Fe film, the following workflow was applied. First, the “dark” signal (laser off, X-ray off) was subtracted from each azimuthally integrated XRD pattern. Next, to correct for shot-to-shot fluctuations of the EuXFEL’s SASE beam, every dark-corrected pattern was normalized by scaling its mean intensity within the $2\theta = 19° - 20°$ range to unity. This angular window was selected because it contains no Fe diffraction peaks, is unaffected by dark-signal artifacts (see Fig. S3a), and is dominated by the SiN/$SiO_2$ scattering peak which exceeds the intensity of liquid or diffuse scattering from Fe. Consequently, it serves as a direct proxy for the incident X-ray flux. Finally, the residual background – constructed from the normalized, laser-off pattern – was subtracted after approximating the Fe(110) Bragg reflection with a linear baseline fit (Fig. S3b).

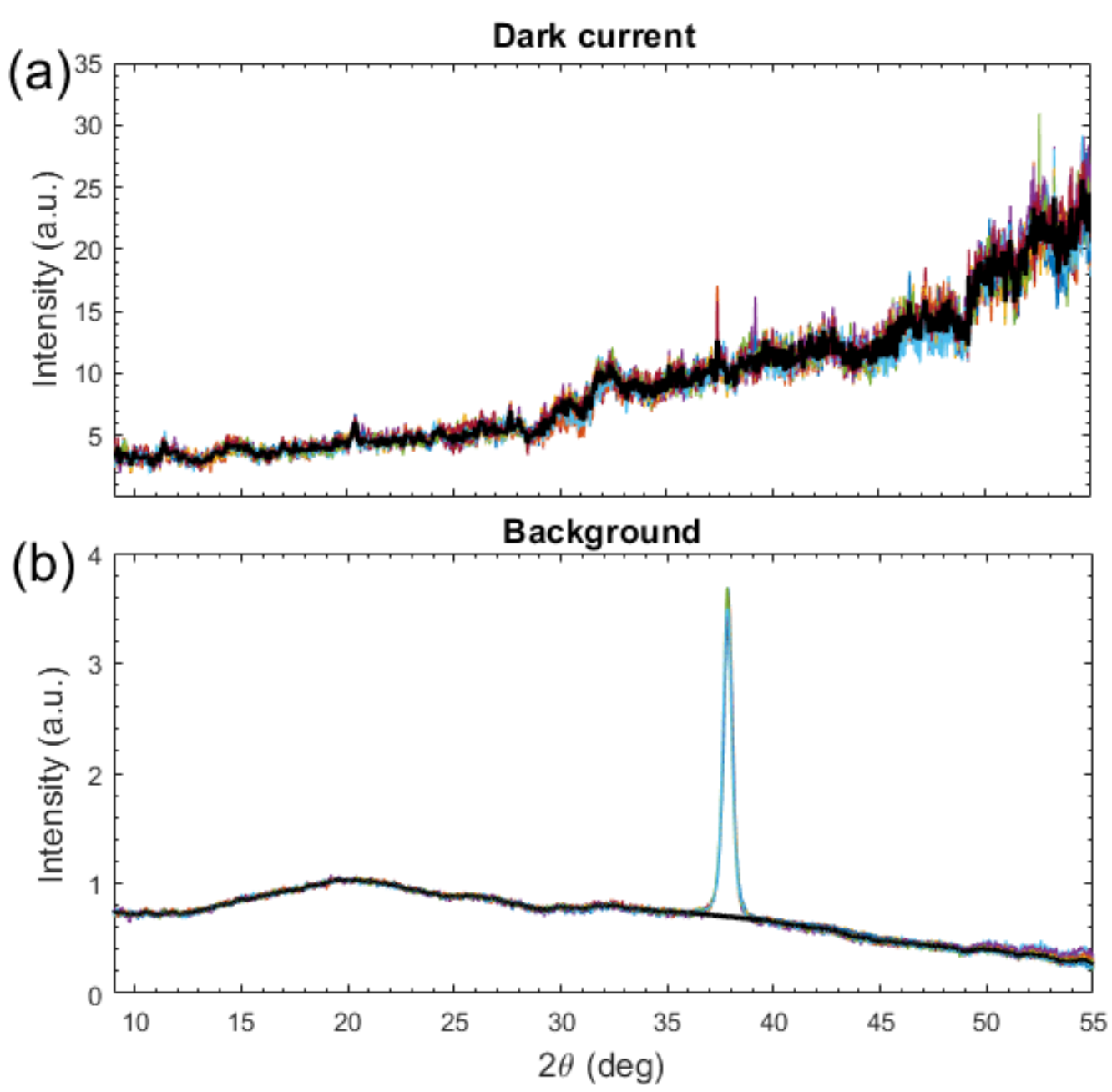

FIG S3. “Dark” background signal (a) and normalized diffraction pattern of the cold sample (b) after “dark” background subtraction which was used for creation of background signal. Both graphs involve multiple

X-ray exposures (different line colors), and thick black lines represent the average curve.

The illustrative example of the above described procedure is shown in Fig. S4.

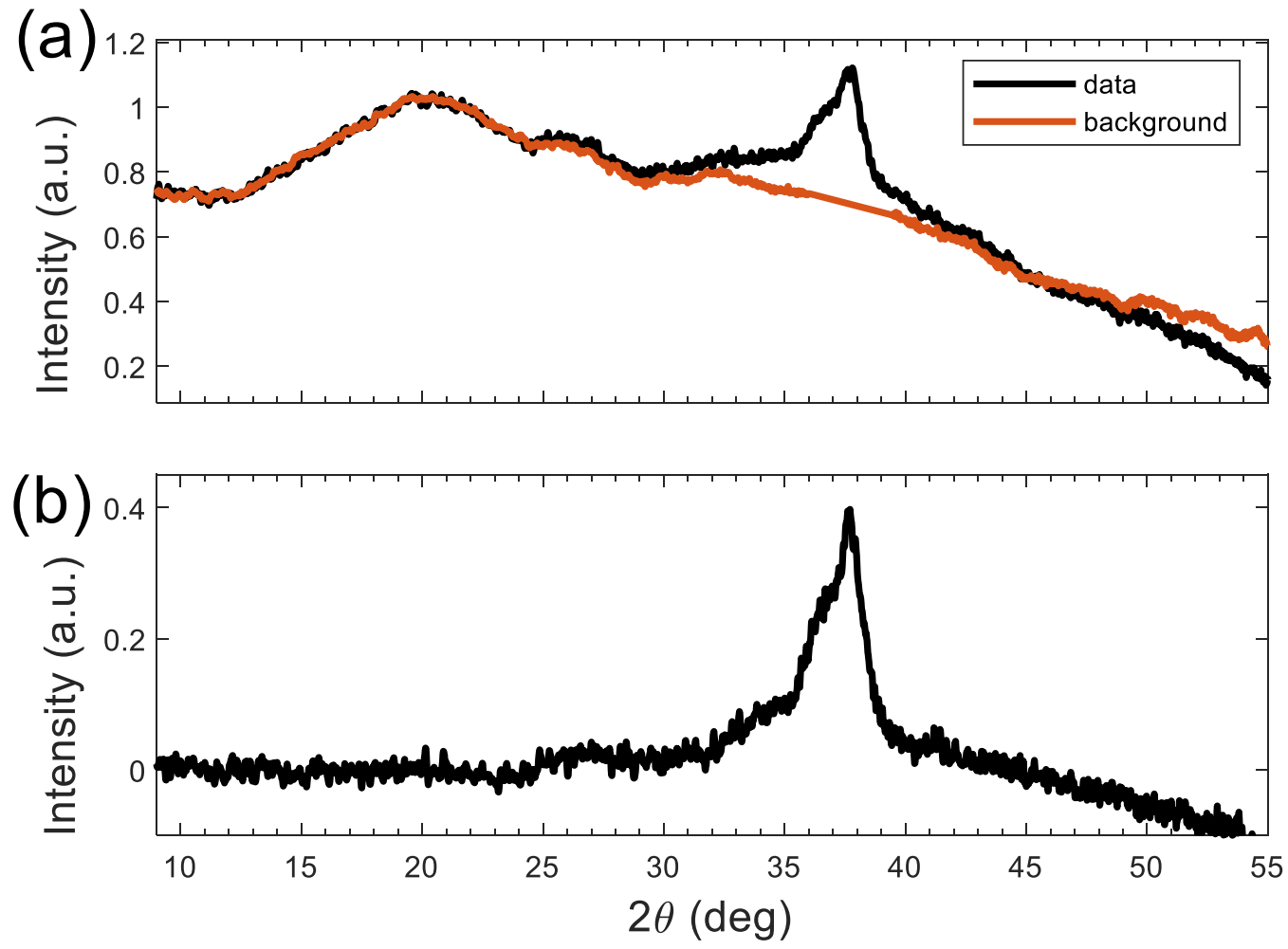

FIG S4. A diffraction pattern of a Fe film (black line, delay time 10 ps and a pump fluence of 166 mJ/cm$^2$), a background signal (red) (a). A diffraction pattern of the Fe film after background subtraction and normalization is shown in (b).

Patterns exhibiting excessive flaws or noise – arising from defective sample windows or unusually low X-ray pulse energies – were excluded from further analysis.

*XRD signal deconvolution into Bragg and liquid peaks*

Peak fitting of the azimuthally-integrated XRD patterns was performed in the 2θ range of $\sim 27° - 45°$, covering the angular region of the (110) Bragg reflections of the Fe bcc crystalline phase or the (110) and $\{(011),(101)\}$ Bragg reflections of the Fe bct crystalline phase and the broad peak originating from the liquid phase. Prior to fitting, a linear background was subtracted from the data in the entire 2θ range $\sim 27° - 45°$, where all peaks including broad "halo" from the liquid are observed. Each fit involved three peak functions. Two Pseudo-Voight (50/50 Gaussian/Lorentzian contribution) functions represented the (110) peak of bcc/bct Fe phase and $\{(011),(101)\}$ peaks of bct-Fe. A Gaussian peak represented the liquid contribution to the diffraction pattern.

The values of peak parameters taken for further analysis were obtained by averaging several measurements taken at equal values of the pump-probe delay time and absorbed energy density.

**Results**

Temporal changes of the fitted peak parameters (integrated intensities – peak areas, positions in momentum transfer q and FWHM) for the case of partial melting of the sample (a: $F_i = 115$ mJ/cm$^2$; $F_i = 191$ mJ/cm$^2$) are shown in Fig. S5. The decrease of the integrated intensities of the Bragg peaks is smaller for lower fluence (Fig. S5a), while the integrated intensity of the liquid peak is bigger for higher fluence (Fig. S5b). The sharp decreasing of the intensity for a pump fluence of 115 mJ/cm$^2$ is due to much higher timestep of 2 ps (comparing with 0.5 ps for pump fluences of 166 and 191 mJ/cm$^2$) which do not allow to capture ultrafast changes in intensities due to heating and melting.

The behavior of Bragg peaks positions is similar to the described in manuscript case of a pump fluence of 166 mJ/cm$^2$ (taking into account higher timestep for pump fluence of 115 mJ/cm$^2$). However, for all fluences the maximum peak splitting due to the shift of the high-q peak towards higher q and of the low-q peak to lower q is observed at delay time of about 12.5 ps, and increase with fluence.

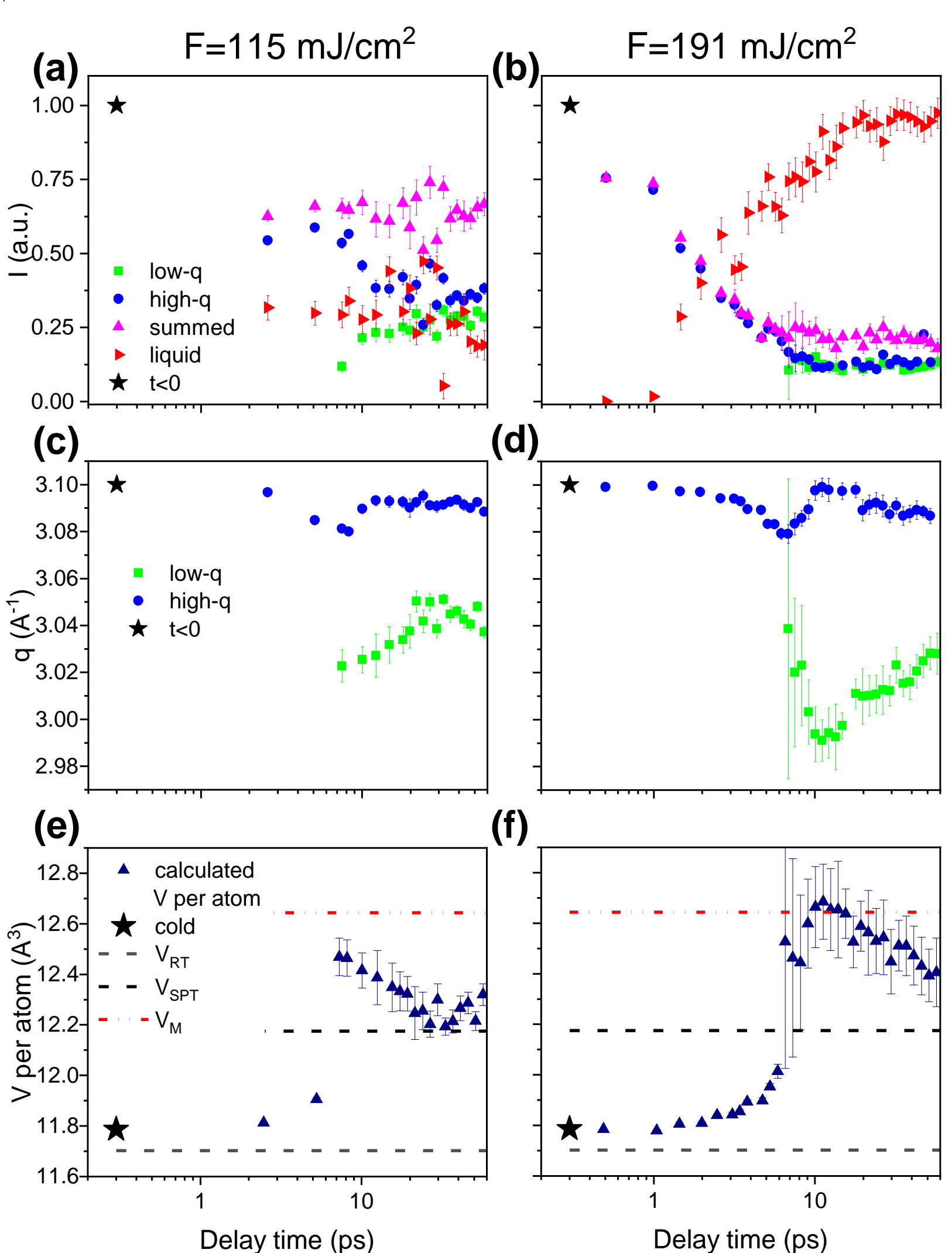

FIG S5 Dependencies of peak parameters as a function of the delay time for the pump fluences of 115 and 191 mJ/cm$^2$: (a, b) integrated intensities normalized to cold value (for solid) or to fully molten Fe layer (for liquid), (c, d) Bragg peak positions, (e, f) volume per atom.

As it is described in the manuscript, if the fcc-Fe phase would form, the summed intensity of the Bragg peaks should decrease by factor of 2 for complete bcc-fcc phase transition. For a pump fluence of 115 mJ/cm$^2$ the region of delays of approx. 6 – 9 ps where the low-q peak appears and its intensity increases, should be examined more closely (Fig. S5a). The low-q peak emerges at approximately 7 ps after irradiation (the time step for this dataset is larger than that for the pump fluence of 166 mJ/cm$^2$ in the main text). The intensity of the low-q peak increases as the intensity of the high-q peak simultaneously decreases for delays of approximately 6 – 9 ps. Clearly, the summed intensity of the Bragg peaks barely changes during this time. This further rules out a bcc-to-fcc phase transition.

The temporal changes in the volume per atom for the pump fluences of 115 and 191 mJ/cm$^2$, as shown in Fig. S5e and S5f, exhibit a similar behavior to the 166 mJ/cm$^2$ case discussed in the main manuscript. The maximum value is attained at a delay time of approx. 12.5 ps (corresponding to the time where the peak splitting is maximum). For $F_i =$ 115 mJ/cm$^2$, $V$ is equal to 12.5 A$^3$, which is slightly smaller than $V_M$. For $F_i = 191$ mJ/cm$^2$ values of the $V$ are close to $V_M$, as was also observed for $F_i = 166$ mJ/cm$^2$. Please note that in the partial melting regime $V$ never exceeds $V_M$, but also does not drop below $V_{SPT}$.

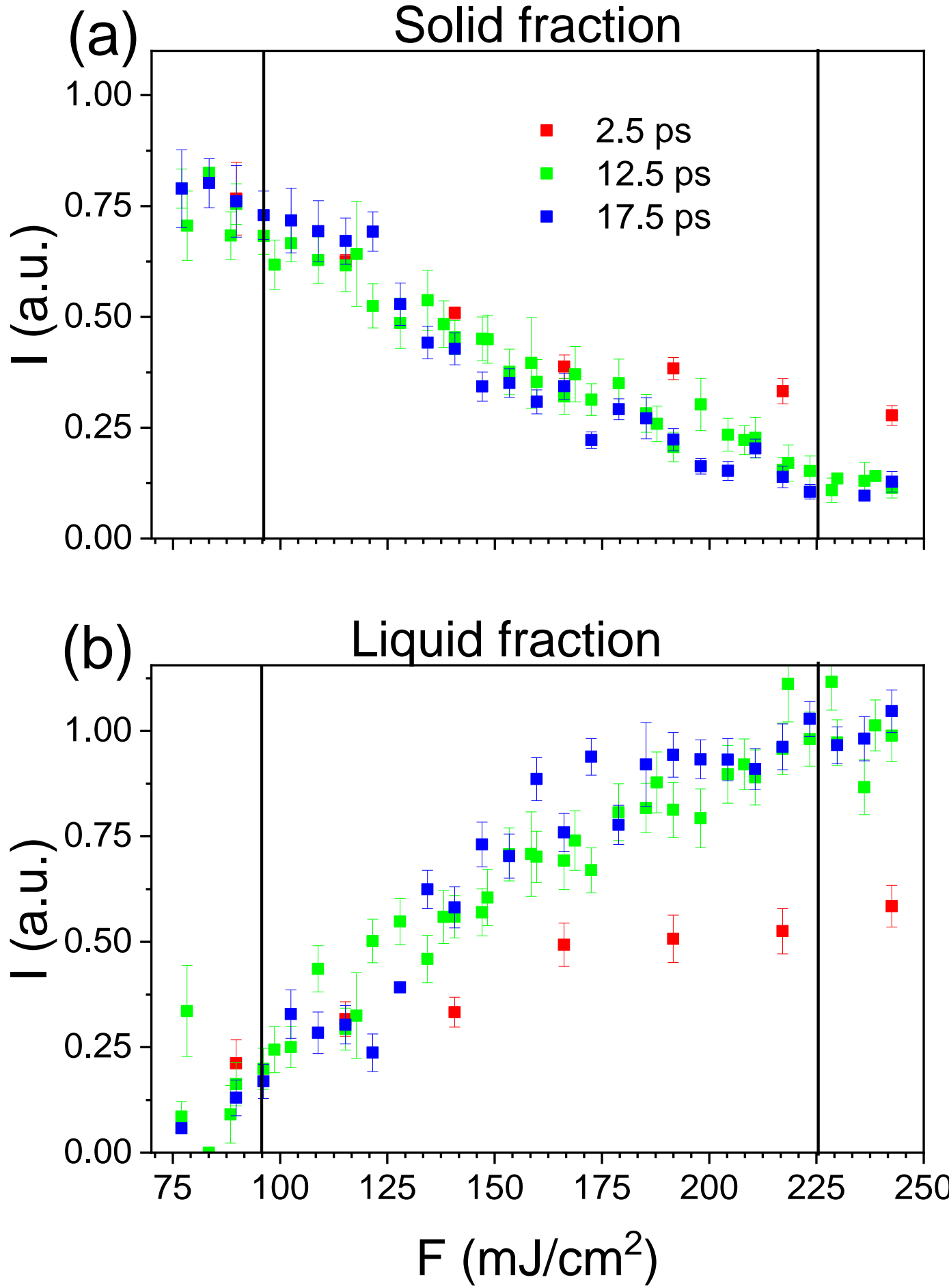

FIG S6 Dependence of the summed integrated intensities of the Bragg peaks normalized to the signal for the as-grown sample (a) and intensity of the liquid peak normalized on the fully molten state (liquid) (b) for 3 delay times: 2.5 (red), 12.5 (green) and 17.5 ps (blue). Vertical lines approximately indicate three regime: heating, partial melting and complete melting.

Fig. S6 presents fluence dependencies of the “solid” (sum of the integrated intensities of the Bragg peaks) and “liquid” fraction (integrated intensity of the Gaussian liquid “halo”) for three different delay times. For a delay time of 12.5 ps (maximum peak splitting and volume per atom), three regimes can be distinguished: no melting for fluences below approx.

95 mJ/cm$^2$, partial melting in the fluence range from 95 to 225 mJ/cm$^2$, and complete melting for $F_i > 225$ mJ/cm$^2$. The observed a small increase of the liquid intensity for fluences below 95 mJ/cm$^2$ we indicate to thermal diffuse scattering originating from the heated solid fraction, similar as in [2].

The volume fraction of both phases doesn't change between 12.5 and 17.5 ps, indicating that the (partial) melting process is completed before 12.5 ps delay time. In the partial melting regime, at the delay time of 12.5 ps when the temperature equilibration over the sample depth is expected, the volume fractions of the crystalline (proportional to the sum of the Bragg peak intensities) and liquid phases (proportional to the broad liquid "halo" intensity) depend non-linearly (decreasing and increasing, respectively) on fluence (Fig. S6).

For pump fluences higher than approx. 225 mJ/cm$^2$, complete melting of the solid phase occurs within a timescale of $3 - 5$ ps, and no further structural changes are observed at later delay times.

## Calculation details

### 3. DW calculations

The decrease in the integrated intensity of the Bragg peaks has been estimated using a parametrized model from the ref. [5]. However, due to the absence of data for the high-temperature δ-Fe phase, parameters from the calculations for α-Fe were used to estimate the Debye-Waller factor for temperature values near to the melting point (approx. 1811K). As a result, a 25% decrease in the Bragg peak intensity compared to the initial "cold" value is expected for the partial melting condition. Similar decrease of the Bragg peak intensity was observed in a recent experimental studies of Pd melting [2].

### 4. Volume per atom calculation

Assuming a three-dimensional (3D) expansion of the lattice, the lattice parameters and the volume per atom were calculated.

The lattice parameter of the bcc phase was calculated using the well-known equation [6]:

$$a_{bcc} = \frac{2\pi}{q}\sqrt{h^2 + k^2 + l^2}$$

where $h$, $k$, $l$ are the Miller indices of the corresponding Bragg reflection (here – (110)) and $q$ – position of the Bragg (low-q or high-q) peak assuming that it originates from the bcc phase.

Using the same equation, the lattice parameter of the fcc phase was calculated:

$$a_{fcc} = \frac{2\pi}{q}\sqrt{h^2 + k^2 + l^2}$$

Here, Miller indices $h$, $k$, $l$ are for (111) Bragg reflection from the fcc phase and q is the low-q peak position.

The lattice parameters of the bct phase $c/a > 1$ case (with low-q peak indexed as (011) reflection and high-q peak as - (110) reflection) were estimated using the following expressions [6] using both low-q and high-q peak positions:

$$a_{bct} = \frac{2\pi}{q_{h-q}}\sqrt{h^2 + k^2 + l^2}$$

$$c_{bct} = \sqrt{l^2/\left(\frac{{q_{l-q}}^2}{4\pi^2} - \frac{(h^2 + k^2)}{{a_{bct}}^2}\right)}$$

where $h$, $k$, $l$ are Miller indices of reflection (110) for $a_{bct}$ and (011) for $c_{bct}$, $q_{h-q}$ and $q_{l-q}$ – are the positions of the high- and low-q peaks, respectively.

In the case of $c/a < 1$, we assume the opposite Bragg peaks indexing, with the low-q peak corresponding to the (110) reflection and the high-q peak corresponding to the (011) reflection. Consequently, we obtained the following equations for lattice parameters:

$$a_{bct} = \frac{2\pi}{q_{l-q}}\sqrt{h^2 + k^2 + l^2}$$

$$c_{bct} = \sqrt{l^2/\left(\frac{{q_{h-q}}^2}{4\pi^2} - \frac{(h^2 + k^2)}{{a_{bct}}^2}\right)}$$

Using these lattice parameters values, the volume per atom for bcc, fcc and bct phases were calculated as:

$$V_{bcc} = \frac{a_{bcc}^3}{2}$$

$$V_{fcc} = \frac{a_{fcc}^3}{4}$$

$$V_{bct} = \frac{a_{bct}^2 \cdot c_{bct}}{2}$$

**Discussion**

**5. Acoustic wave propagation**

Oscillatory behavior caused by acoustic strain waves traveling back and forth in the Fe film isn't observed due to the high overall thickness of the $Si_3N_4$/Fe/$SiO_2$ layers stack. The acoustic round-trip-time for the whole layer stack is approx. 130 ps, which is much higher than the longest delay of 60 ps considered in this work. According to the experimental data, the tetragonal distortion has a much larger influence than any possible uniaxial deformation. For our nanocrystalline Fe layer with randomly oriented crystallites, the 1D strain should shift all Bragg peaks in the same direction [7], while we observe that the low-q and high-q peaks move in opposite directions, as is attributed to the tetragonal distortion effect.

**6. Thermal stress and thermal expansion of the Fe layer**

Using the methodic developed in Refs. [8,9] the influence of mixed stress conditions (a 2D thermal stress in the in-plane direction and 1D uniaxial stress in the out-of-plane direction) on the Fe unit-cell was investigated. As initial conditions, we assumed a thermally expanded lattice (with a lattice parameter of $a = 2.935$ Å near the melting point [10]), elastic constants for δ-Fe near the melting point obtained from theoretical calculations [11], and a random distribution of crystallite orientations in the Fe polycrystalline layer. Changes in the interplanar spacing of the $\{110\}$ plane family and corresponding Bragg peak positions were calculated for the randomly oriented polycrystalline material.

The obtained results indicate that the thermal stress due to ultrafast heating (assuming isotropic conditions just after lattice heating and the formation of the δ-Fe phase due to ultrafast lattice heating) is near 10 GPa in the first few picoseconds after laser irradiation. Furthermore, stress-relaxation in the out-of-plane direction, leads to formation of two maxima in the diffraction patterns. It resembles the situation with the peak splitting observed in the experimental data. However, to reconstruct the experimental pattern at approx. 12.5 ps (maximum peaks splitting), the stress values should be -1.65 GPa for the out-of-plane direction and 0.35 GPa in the in-plane direction – which is unrealistic. It indicates that the uniaxial stress conditions cannot explain the splitting of the Bragg peaks.

**References**

[1] D. Khakhulin et al., *Ultrafast X-Ray Photochemistry at European XFEL: Capabilities of the Femtosecond X-Ray Experiments (FXE) Instrument*, Appl. Sci. **10**, 995 (2020).

[2] J. Antonowicz et al., *Structural Pathways for Ultrafast Melting of Optically Excited Thin Polycrystalline Palladium Films*, Acta Mater. **276**, 120043 (2024).

[3] J. Chalupský et al., *Spot Size Characterization of Focused Non-Gaussian X-Ray Laser Beams*, Opt. Express **18**, 27836 (2010).

[4] G. Ashiotis, A. Deschildre, Z. Nawaz, J. P. Wright, D. Karkoulis, F. E. Picca, and J. Kieffer, *The Fast Azimuthal Integration Python Library: PyFAI*, J. Appl. Crystallogr. **48**, 510 (2015).

[5] H. X. Gao and L.-M. Peng, *Parameterization of the Temperature Dependence of the Debye–Waller Factors*, Acta Crystallogr. Sect. A Found. Crystallogr. **55**, 926 (1999).

[6] B. D. Cullty and S. R. Stock, Elements of X-Ray Diffraction: Third Edition (Prentice-Hall, 2001).

[7] W. Lu, M. Nicoul, U. Shymanovich, F. Brinks, M. Afshari, A. Tarasevitch, D. Von Der Linde, and K. Sokolowski-Tinten, *Acoustic Response of a Laser-Excited Polycrystalline Au-Film Studied by Ultrafast Debye-Scherrer Diffraction at a Table-Top Short-Pulse x-Ray Source*, AIP Adv. **10**, 35015 (2020).

[8] A. Yoneda and A. Kubo, *Simultaneous Determination of Mean Pressure and Deviatoric Stress Based on Numerical Tensor Analysis: A Case Study for Polycrystalline x-Ray Diffraction of Gold Enclosed in a Methanol–Ethanol Mixture*, J. Phys. Condens. Matter **18**, S979 (2006).

[9] M. Gauthier, *Effects of Uniaxial Stress on X-Ray Diffraction Spectra*, High Press. Res. **22**, 779 (2002).

[10] Z. S. Basinski, W. Hume-Rothery, and A. L. Sutton, *The Lattice Expansion of Iron*, Proc. R. Soc. London. Ser. A. Math. Phys. Sci. **229**, 459 (1955).

[11] J. Neuhaus, M. Leitner, K. Nicolaus, W. Petry, B. Hennion, and A. Hiess, *Role of Vibrational Entropy in the Stabilization of the High-Temperature Phases of Iron*, Phys. Rev. B **89**, 184302 (2014).